\documentclass[conference]{IEEEtran}
\usepackage{amsthm}

\usepackage{amssymb}

 \usepackage{cite}
\usepackage[utf8]{inputenc}
\usepackage[english]{babel}
\usepackage{csquotes}

\ifCLASSINFOpdf
\usepackage[pdftex]{graphicx}
\else
\fi

\usepackage[cmex10]{amsmath}
\usepackage[ruled,vlined,linesnumbered]{algorithm2e}
\SetKwFor{For}{for }{ $\lbrace$}{$\rbrace$}

\SetCommentSty{mycommfont}
\usepackage{array}
\usepackage{url}
\usepackage{mathrsfs}

\usepackage[dvipsnames,svgnames]{xcolor}
\usepackage{bbm}

\begin{document}

\title{Impact of COVID-19 on Mobility and Electric Vehicle Charging Load}

\author{\IEEEauthorblockN{Alejandro Palomino\IEEEauthorrefmark{1}, Masood Parvania\IEEEauthorrefmark{1}, Regan Zane\IEEEauthorrefmark{2}}
\IEEEauthorblockA{\IEEEauthorrefmark{1}Department of Electrical and Computer Engineering, University of Utah, Salt Lake City, UT 84112
\\
\IEEEauthorrefmark{2}Department of Electrical and Computer Engineering, Utah State University, Logan, UT 84322}
Emails: \{alejandro.palomino, masood.parvania\}@utah.edu, regan.zane@usu.edu}

\maketitle
\begin{abstract}
The COVID-19 pandemic has depressed overall mobility across the country. 
The changes seen reflect responses to new COVID-19 cases, local health guidelines, and seasonality, making the relationship between mobility and COVID-19 unique from region to region.
This paper presents a data-driven case study of electric vehicle (EV) charging and mobility in the wake of COVID-19. 
The study shows that the number of EV charging sessions and total energy consumed per day dropped by 40\% immediately after the arrival of the first COVID-19 case in Utah.
By contrast, the energy consumed per charging session fell by just 8\% over the same periods and the distribution of session start and end times remained consistent throughout the year.
While EV mobility dropped more dramatically than total vehicle mobility during the first wave of COVID-19 cases, and returned more slowly, both returned to stable levels near their mean values by September 2020, despite a dramatic third wave in new infections.
\end{abstract}


\section{Introduction}
Nearly one year after the first case of COVID-19 was reported in the United States on January 20th, 2020, mobility remains depressed as travelers seek to avoid the risk of COVID-19 transmission. 
Nationally, visits to retail and recreation, transit stations, and workplaces remain 14\%, 32\%, and 30\% below baseline, respectively \cite{google2020mobility}. 
Mobility and COVID-19 transmission are positively correlated across the country, but changes in infections lag those in mobility by up to three weeks \cite{badr2020association}. 
This lag challenges effective policy development to curb the COVID-19 transmission. 

In response to COVID-19 precautions and stay-at-home recommendations from health experts, corporations and their employees have embraced work-from-home opportunities.
Seventy-seven percent of employees now work from home as compared to just 9\% prior to the COVID-19 outbreak \cite{analytics2020global}.
This shift reflects a 68\% increase in employees who now work from home 5 days per week.
As more employees work from home fewer get into vehicles for their daily work commutes, which comprise 682 billion, or 28\%, of annual vehicle miles traveled in the U.S.~\cite{nhts2017}. 

The overall decline in mobility offers an opportunity to explore how this shift manifests in electric vehicles (EVs) as distinct from internal combustion engine vehicles (ICVs).
Public transportation, which may expose users to other people, exhibits decreases in utilization that exceed the drop in mobility broadly. 
New York City, one of the most transit-dense cities in the country, observed a 92\% decrease in public transit ridership during the heart of the infection wave in April 2020 \cite{joselow2020littlerisk}. 
Ride-hailing trips from Transportation Network Providers such as Uber and Lyft decreased by up to 80\% in the spring of 2020 when compared to the same period in 2019 \cite{du2020ridehailing}.
In some instances public transit, car pooling, and micro-mobility providers have limited or suspended service.
These shared transit modes were, and continue to be, supplanted by private car travel. 

Although the private (or personal-use) car market is dominated by ICVs, EVs represent an increasing share. 
This paper explores how private car travel has changed between these two vehicle classes in response to COVID-19. 
Additionally, this work explores how EV charging behaviors and energy demands have changed since the onset of the pandemic.
Typically, EV driving patterns differ slightly from ICV driving patterns due to their relatively limited range, sparseness of charging stations, and time required to recharge. 
Industry, utility, and local government stakeholders across the country recognize the persistence of range anxiety as a barrier to EV adoption and have advanced measures for the deployment of EV charging infrastructure ultimately decreasing air pollution and increasing electricity sales.  
Further energizing this work is the positive correlation observed between publicly available EV charging infrastructure and EV adoption \cite{palomino2019advanced}.

To the authors' knowledge, this study presents the first exploration of EV mobility in comparison to ICV mobility in the time of COVID-19.
The analysis presented relies upon national mobility trends with a focus on available traffic volume and EV charging data in the state of Utah. 
Section \ref{sec:Method} reviews the available data, scope of analysis, and sets the context for the results presented in Section \ref{sec:Results}.
High-level implications of these findings are presented Section \ref{sec:Conclusion}.

\section{Data and Methodology}\label{sec:Method}
This paper identified four sources of data, in addition to COVID-19 case numbers \cite{nytCOVID}, for this data-driven analysis of EV and ICV driving. 
Mobility data lays the groundwork for private car travel in the United States and Utah.
Utah traffic volume data offers highway specific transit patterns.
When observed in before, during, and after temporal contexts, Utah EV charging data can be used to study changes in EV driving. 
Mobility and traffic data is set in time by overlaying COVID-19 event data in Utah. 

\subsection{Mobility Data}
This work employs Google COVID-19 Community Mobility Reports (CMRs) to inform mobility analysis \cite{google2020mobility}.
The study of mobility is incorporated to understand the dominant trends that exist in the presence of holidays, days of the week, and seasonal changes.
This high-level depiction of mobility allows for separation and observation of COVID-19 impacts on human mobility. 
The CMRs are taken from anonymized Google location data and track changes in destination mobility.
Google has categorized all destinations as one of \textit{retail and recreation}, \textit{grocery and pharmacy}, \textit{parks}, \textit{transit stations}, \textit{workplaces}, or \textit{residential}.
Formally, the five weeks of January, an approximate ``pre-COVID" period in the United States, are used as a baseline for mobility comparison. 
Note, CMR data does not offer multi-year comparison data or feature correction to allow for controlled comparison of mobility throughout the COVID-19 pandemic. 
Baselining is limited to the five week period spanning from January 3 to February 6, 2020. 
Percent mobility changes are measured in comparison to this period for the corresponding day of the week.  

\subsection{Utah Traffic Volume Data}
Highway sensors around Utah record passing vehicles by anonymously sensing on-board electronic signals such as bluetooth and Wi-Fi \cite{udot2020covidData}. 
A vehicle picked up by two sensors is called a ``match."
Matches do not provide a precise measure of traffic volume, but inform average speed and relative changes in vehicle movement.
A primary highway in the Salt Lake City metro area is selected as representative of metro-area traffic throughout the state.
The Nov. 2019 - Nov. 2020 dataset for this stretch of highway includes over 1.5 million vehicle matches.
Focus on the Salt Lake City metro area follows from the increased risk of COVID-19 transmission in urban areas, social distancing and health guidelines non-withstanding, and the fact that 31\% of the state of Utah lives in Salt Lake County. 

\subsection{Utah EV Charging Data}
Session level EV charging data, charger identity, charger location, session start time, session end time, and energy, is taken from networked EV charging stations around the state comprising more than 28,000 charging sessions, 2,000 EV drivers, and 300 MWh of energy over the Nov. 2019 to Nov. 2020 analysis period \cite{chargepoint2020Data}. 
Session level data informs EV charging patterns before, during, and since the onset of COVID infections. 
This work analyzes local and long distance driving modes by grouping chargers into public, highway, or other classes.
Public chargers are those located at commercial businesses, public parks, and municipal buildings along surface streets in cities and suburbs. 
The public charger dataset includes 18,000 sessions across 20 level 2 (L2) chargers. 
Highway chargers are located at gas stations along major highways in support of long distance EV trips.
The highway charger dataset includes 800 DC fast and 400 L2 charging sessions across a total population of 8 DC fast and 8 L2 chargers. 

\subsection{Utah COVID-19 Event Data}
The trends depicted by mobility, traffic, and EV charging data are set in time by key events, such as case spikes and health guidelines, during COVID-19 in Utah.
The occurrence of these events is taken from the Utah COVID-19 response team's timeline \cite{utahCOVIDevents}.
These events are overlayed on top of mobility trends to better understand how the public at large responded to COVID-19 and focus on how those responses manifest among ICV and EV drivers. 
On February 28th, the first COVID-19 patient arrived in Utah. 
In response, the state government issued the ``Stay Safe, Stay Home" order on March 27th.
On April 17th, the state announced its plan for re-opening. 
The 49 days between the arrival of the first case and announcement of re-opening correspond with the greatest change in mobility. 

\section{Analysis and Discussion}\label{sec:Results}
\begin{figure}[ht]
  \centering
  \vspace{-5pt}
    \includegraphics[width=1\columnwidth]{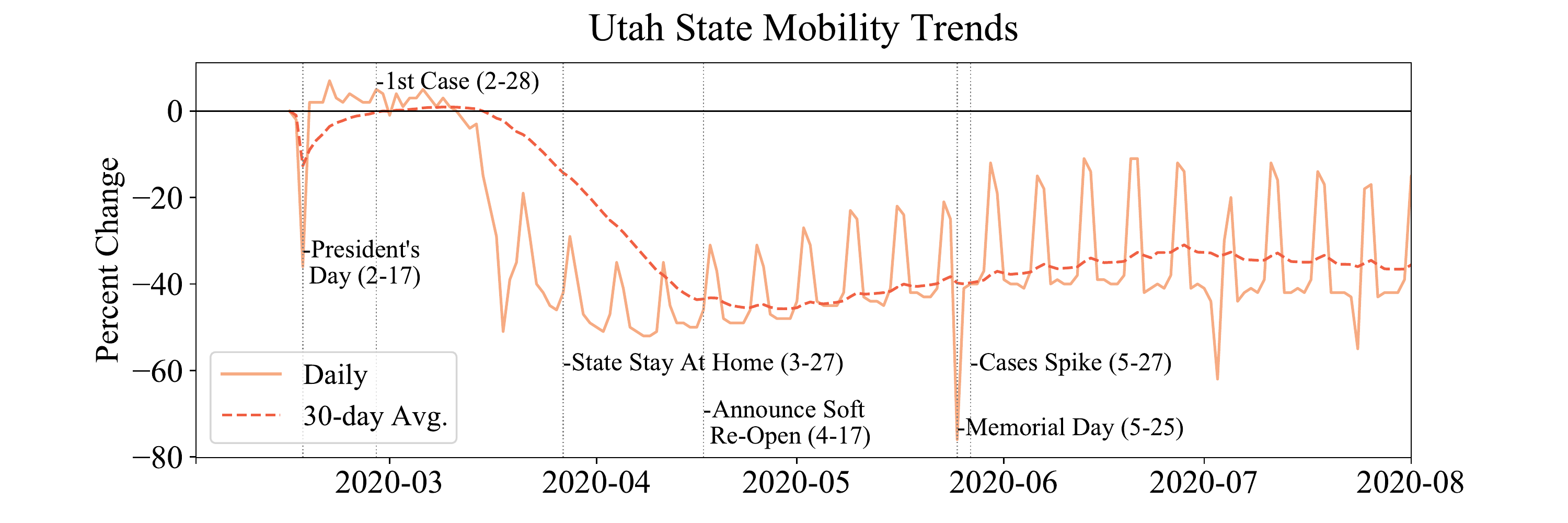}
    \caption{Daily percentage change in workplace visits across Utah.}
    \label{fig:mobilityWork}
  \vspace{-5pt}
\end{figure}

Rolling averages offer a means to distinguish the COVID-19 influence on mobility from normal daily patterns.
Changes in COVID-19 cases and consequent health guidelines have long-term, greater than three weeks \cite{badr2020association}, impacts on mobility in aggregate. 
This study considers day-to-day oscillations in mobility, or more precisely changes in mobility by type of location visited, as signal noise that is short in period and counter-indicative of long-term COVID-19 caused trends. 
Daily workplace visitation, as measured by changes in mobility, depicted in Fig. \ref{fig:mobilityWork}, illustrates the difference in short and long term mobility trends.
Here, the regular oscillations of transit patterns are present in the daily values.

The percent change in workplace mobility first drops due to the President's Day holiday and then remains close to zero percent indicating a continuation of regular mobility patterns, as compared to the baseline period in January. 
The daily results presented in Fig. \ref{fig:mobilityWork} depict normal, weekly oscillations. 
Regular, positive weekend spikes in workplace mobility represent a regression to the baseline mobility as fewer workplace commutes occur on weekends. 
A large decrease in workplace mobility occurs on the Memorial Day holiday followed by a short-term doubling in the number of new daily COVID-19 cases \cite{nytCOVID}.
The close occurrence of the holiday weekend and an increase in COVID-19 is an example that fits into the literature establishing a correlation between mobility and COVID-19 transmission \cite{badr2020association}.
Rolling averages offer a means to filter out short-term oscillations in mobility, expose long-term changes, and extract COVID-19 impacts from this data.

\subsection{Mobility}
\begin{figure*}
  \centering
  \vspace{-5pt}
    \includegraphics[width=1\textwidth]{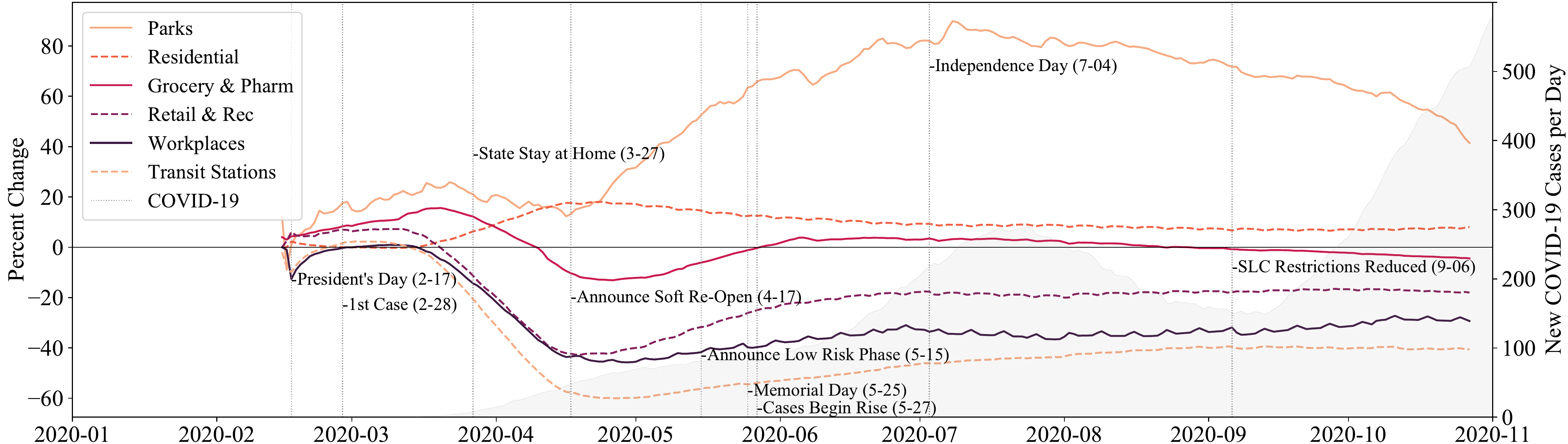}
    \caption{Salt Lake City area mobility timeline over a 30-day rolling average.}
    \label{fig:mobilitySLC}
  \vspace{-5pt}
\end{figure*}

Changes in mobility by destination classification is presented in Fig. \ref{fig:mobilitySLC} according to Google Community Mobility Reports (CMRs) data.
The 30-day rolling average captures both long-term COVID-19 changes in mobility as well as seasonal changes.
A significant increase in park visitation in the Salt Lake City area occurs in the spring and summer.
Throughout the western United States, increased park visitation during the spread of COVID-19 was driven by the change in season \cite{rice2020understanding}.
Recall that the baseline for mobility is January, 2020 and CMRs do not offer year-over-year analysis, so it is expected that park visitation greatly increase in the summer as compared to a winter baseline. 
%

Controlling for short term oscillations and seasonal change clarifies the connections between mobility, COVID-19 transmission, government policies, and the public's response.
A decrease in visitation to public locations beginning in March, reaching their lowest levels in April, are reflective of the first wave of COVID-19 cases in Salt Lake City.
The sustained increase in visitation to residential locations suggests people have shifted recreational trips from public locations to private residences they consider safer. 
Workplace mobility remains far below baseline levels as employers adopt work-from-home policies \cite{analytics2020global} and illustrates the subtle weekly periodicity of the dominant Monday to Friday work week.
Only grocery and pharmacy mobility has returned to near baseline levels indicating the regular need for food and medicine for a population regardless of the season, holiday or COVID-19 events. 
Although still depressed, mobility trends in the Salt Lake City area have stabilized and disconnected from the trajectory of new COVID-19 case numbers since a marked decrease during the first wave of cases in April and May.  
 
\subsection{EV Charging Preferences}
The magnitude and timing of EV charging demand are important to utility load planning \cite{palomino2020data,palomino2020Bayesian}. 
This work offers a unique insight into how EV charging demand profiles change during a pandemic. 
COVID-19 infection trends are classified into three primary time periods, ``before," ``1st wave," and ``1st fade," according to their relation to the 49 days between the state's first case on February 28th and the state's announcement on April 17th of an economic recovery plan for reopening \cite{utahCOVIDevents}.
Within the 1st wave period, the state issued a ``Stay Safe, Stay Home" directive on March 27th in response to the growing number of new COVID-19 cases. 
The 49 days immediately after the announcement of the reopening plan represents the 1st fade analysis period.
Since April 17th, the period since the directive, Salt Lake City has experienced two additional waves of new COVID-19 cases.

The 1st wave period correlates with the biggest change in EV charger utilization.
Average daily energy demand from all EV chargers in Salt Lake City dropped from approximately 1200 to 700 kWh per day in the weeks immediately before and after COVID-19 in Utah, as shown in Table \ref{tbl:aggEV}. 
The number of sessions per day and total EV charging energy dropped by 35\% and 40\% respectively, but charging energy per session dropped by just 8\%.
This discrepancy indicates that despite a large change in charger utilization in the aggregate, individual charging session preferences remained fairly constant.
Similarly, a modest decrease in active charging duration of 12\% occurred. 
The drop in active charging duration and session energy may suggest that EV drivers took shorter trips during the ``Stay Safe, Stay Home" period and thus had lower EV range demands, but EV trip specific data is not available to make that claim explicitly.
The 1st fade period exhibits a slight rebound in session level energy, but otherwise characterizes a decrease in total non-residential EV charger utilization.
\begin{table}[h]
  \centering
  \vspace{-5pt}
    \caption{Aggregate EV Charger Utilization Measures}
    \label{tbl:aggEV}
    \includegraphics[width=1\columnwidth]{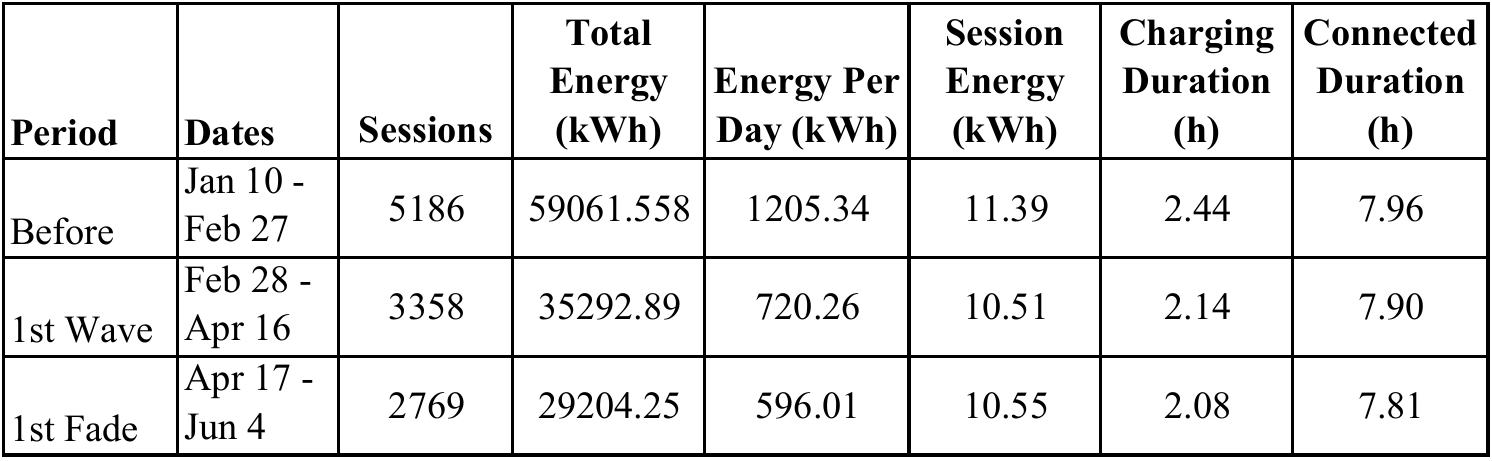}
    \vspace{-5pt}
\end{table}

\begin{figure}[ht]
  \centering
    \includegraphics[width=0.9\columnwidth]{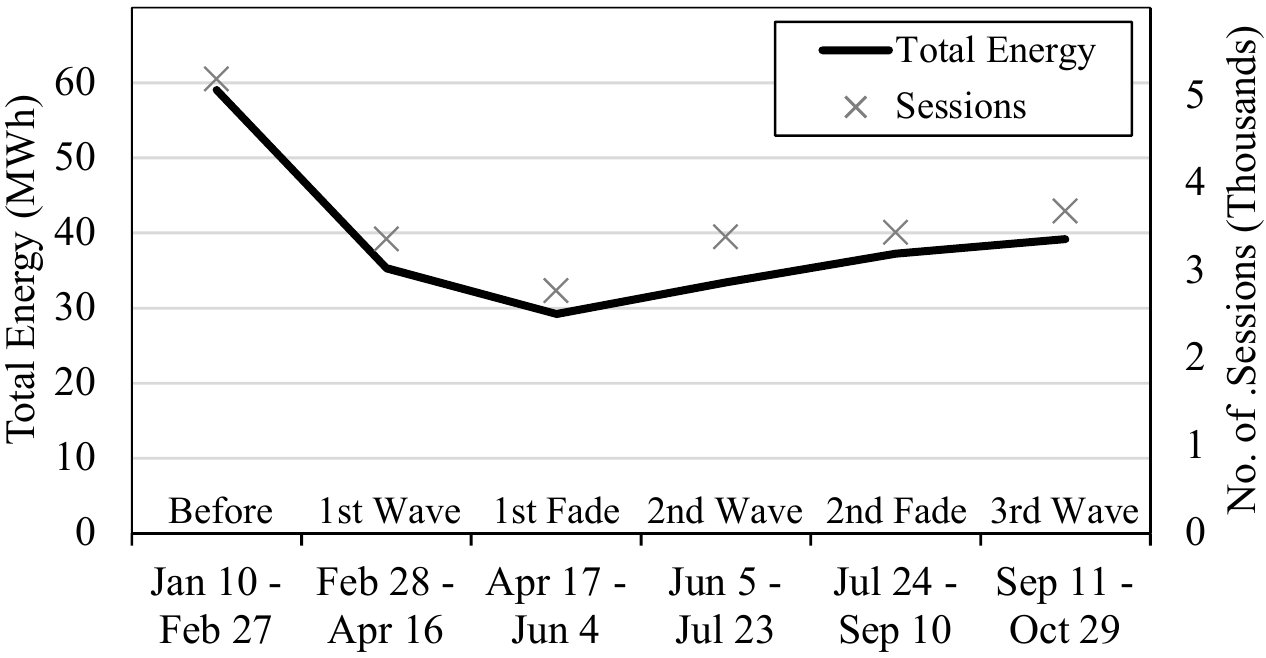}
    \caption{Total EV charging energy and number of sessions.}
    \label{fig:slcTotals}
\end{figure}

Beyond the ``1st fade" period from April 17th to June 4th, this study classifies three additional 49 day periods, ``2nd wave," ``2nd fade," and ``3rd wave" to explore the longer term recovery of EV charging after the initial onset of COVID-19 transmission and statewide response. 
These periods span the dates of June 5th - July 23rd, July 24th - September 10th, and September 11th - October 29th. 
These subsequent periods continue the modest rebound in total EV charger utilization after the ``Stay Safe, Stay Home" directive. 
This trend is depicted in the total energy and number of sessions plotted per period in Fig. \ref{fig:slcTotals}.
The lowest period of EV charger utilization correlates with the period of lowest mobility shown in Fig.~\ref{fig:mobilitySLC}.
Over these six periods, average energy consumption per session ranged between 9.89 and 11.39 kWh.

\begin{figure}[ht]
  \centering
    \includegraphics[width=1\columnwidth]{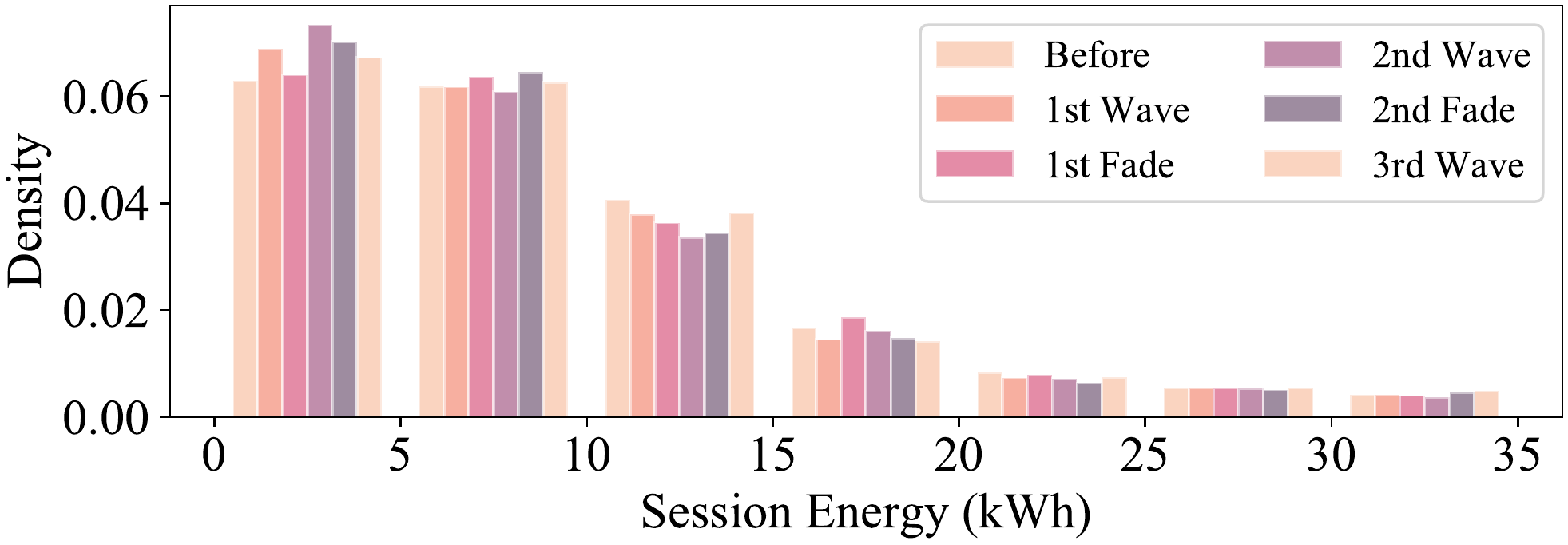}
  \vspace{-15pt}
    \caption{Distribution of EV charging session energy during COVID-19.}
    \label{fig:seshEnergyHist}
  \vspace{-10pt}
\end{figure}

The consistency of per session energy consumption is further underscored by the comparison histogram in Fig. \ref{fig:seshEnergyHist}.
The distribution of sessions by energy consumption remains consistent throughout all six periods outlined in this work. 
Most EV charging sessions consume less than 10 kWh per session, corresponding to an added EV range of up to 30 miles \cite{epa2011new}.
Comparing the added 30 mile range to the average U.S. vehicle trip length of 10.5 miles illustrates the capabilities of EVs to perform similarly to ICV for most trips \cite{nhts2017}. 

\begin{figure}
  \centering
  \vspace{-5pt}
    \includegraphics[width=1\columnwidth]{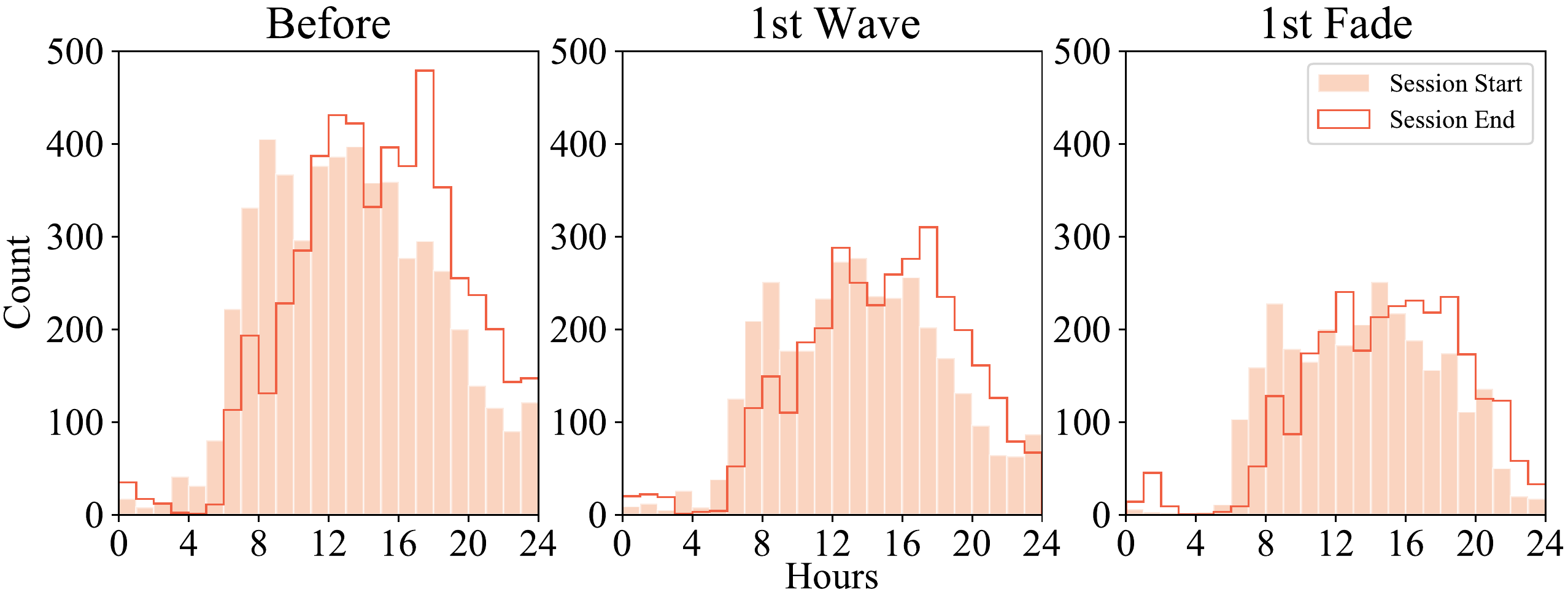}
  \vspace{-15pt}
    \caption{Distribution of EV charging session start and end times during
    COVID-19.}
    \label{fig:sessionTimeHist}
\end{figure}

Charging session preferences can be further expressed according to session start and session end time in Fig. \ref{fig:sessionTimeHist}. 
Session start time for charging at publicly sited EV chargers follows a bi-model distribution which corresponds with EV commuters arrival to work and return from lunch  \cite{palomino2019advanced}.
The broad adoption of remote work and school practices does not change the dominant EV charging session time behaviors, but makes them less prominent.

The total number of EV charging sessions in Salt Lake City dropped significantly after the arrival of COVID-19, but the profile shape remains consistent in Fig. \ref{fig:sessionTimeHist}. 
The decrease in the number of sessions overall in the 1st wave and 1st fade periods nearly eliminates charging sessions late in the evening or early in the morning.
These lost sessions correlate with public locations associated with retail and recreation where social distancing may eliminate the kind of long visits that would accommodate public charging.

\subsection{EV Charger Utilization}
Publicly available EV charger energy consumption, a surrogate for EV driving, offers an indication of EV driving trends in comparison to highway traffic volume data.  
This study focuses on public and highway EV chargers to illustrate how mobility patterns changed during COVID-19 regarding local and long distance EV travel. 
Recall, public chargers in the studied dataset are located along surface streets in near cities and highway chargers are located along highways. 
\begin{figure*}
  \centering
  \vspace{-5pt}
    \includegraphics[width=1\textwidth]{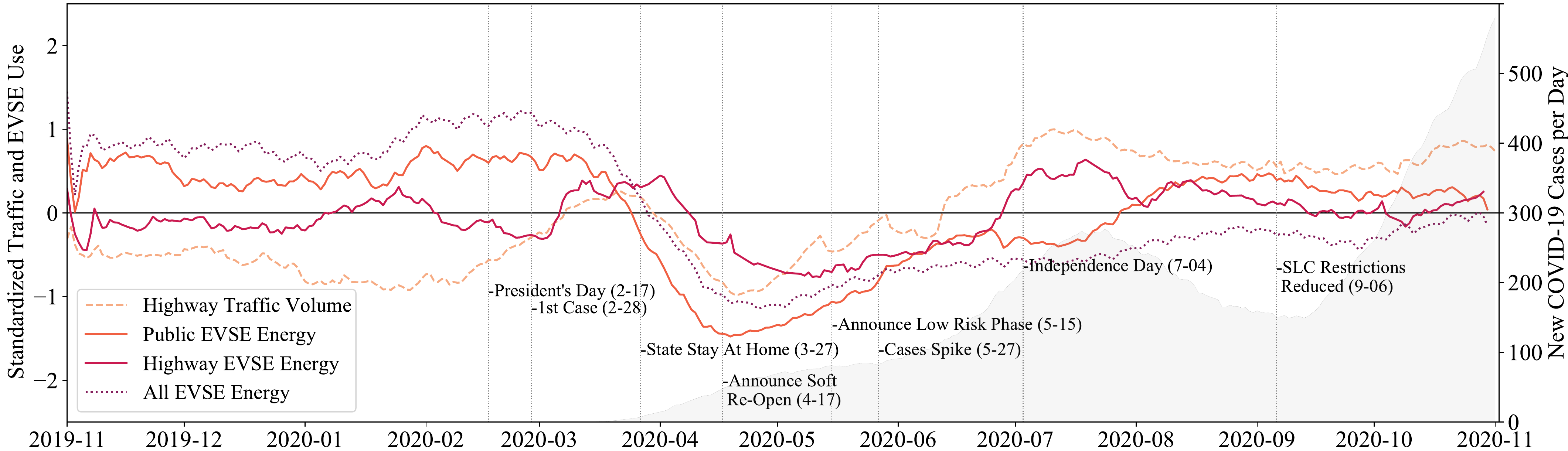}
    \caption{Utah highway traffic and electric vehicle supply equipment (EVSE) utilization timeline (30-day rolling average).}
    \label{fig:timeline}
  \vspace{-5pt}
\end{figure*}

The Utah traffic volume and EV charging data are standardized in Fig. \ref{fig:timeline} for side-by-side comparison. 
Highway traffic volume data comprises travel by both ICVs and EVs, but mostly represents ICVs as EVs comprise a small percentage of the vehicle market \cite{palomino2019advanced}.
Highway traffic exhibits a seasonal trend with increased volumes in the summer.
The results in Fig.~\ref{fig:timeline} shows the start of increased traffic in the spring, followed by a COVID-19 disruption, and finally slightly increased summer travel. 
Collectively, both EV and total highway traffic respond to the state's ``Stay Safe, Stay Home" directive for social distancing on March 27th.
Public EV chargers most support local EV commuting for trips to work, errands, and recreation.
These public EV chargers are concentrated in urban centers where COVID-19 infection has increased the fastest.
Consequently, public EV chargers energy exhibits the clearest drop in response to the arrival of COVID-19. 
EV drivers tend to come from high income households with high educational attainment and highly educated employees are most likely to work from home \cite{blsWorkfromHome, farkas2018environmental}.
This data suggests that EV drivers are more likely to work from home than the average employee resulting in a deeper mobility drop for EV drivers. 

The trajectory of public EV charger utilization and highway traffic volume changed on the date that the State of Utah announced its plans to re-open its economy, but while highway traffic volume returned to its mean value soon after the Memorial Day holiday, public EV utilization lagged behind indicating that a smaller share of EVs were on the road after the emergence of COVID-19. 
The lagging reaction of public EV charger utilization to the 2nd wave and 3rd wave periods of COVID-19 infection suggest a pattern in which EV drivers increasingly return to the road as the number of new COVID-19 cases slows down and then temporarily reduce their driving when cases accelerate.
It remains to be seen if the latest rise in cases will similarly depress EV driving.
This phenomenon is not observed in the highway EV charger utilization trend and offers a distinction between EV and ICV driving preferences. 

Highway EV charger utilization presents a more erratic trend.
The highway EV chargers available are sparsely distributed across the state where the spread of COVID-19 has varied from community to community. 
For example, while the latest count for new COVID-19 cases in Salt Lake City region was 800, there were just 28 new cases the neighboring Tooele county \cite{nytCOVID}. 
The varied exposure of different regions of the state to the pandemic results in diverse responses in mobility which may account for the erratic utilization trend exhibited by highway chargers.
As of approximately Sept. 2020, public and highway EV charger utilization returned to its mean value as measured on a year-over-year basis while highway traffic continues to present increased volume after the summer travel season. 
Conversely, EV utilization measured across all EV chargers continues to lag behind its mean value.
This discrepancy may be attributeable to the inclusion of workplace chargers, excluded by the public and highway trends, whose utilization is depressed by the COVID-19 reduction in workplace commuters. 
\section{Conclusion}\label{sec:Conclusion}
This analysis shows EV drivers have responded differently than ICV drivers to the pandemic, but individual's EV charging preferences remain consistent over this period. 
Total EV energy consumption dropped by 40 to 50\% during the six COVID-19 periods identified due to a decrease in EV mobility, but session level energy changed by just 8\%. 
EV mobility lagged behind highway traffic volume in recovery after the 1st wave of infections.
Critically, highway traffic volume in the state of Utah has detached from the number of new COVID-19 cases, as indicated by mobility during the 2nd and 3rd waves, implying driver fatigue with COVID-19 risks and stay-at-home recommendations. 
Public EV chargers use, on the other hand, suggests a slight, lagging response of EV drivers to increasing risks during these later infection waves.
When COVID-19 cases increase, EV drivers are more likely than ICV drivers to decrease their mobility and return to driving more slowly once cases decrease.
\bibliographystyle{IEEEtran}
\bibliography{ref}
\end{document}